\documentclass[5p]{elsarticle}

\usepackage{epsfig}

\begin{document}
\title{Morphological diagram of diffusion driven aggregate growth in plane:
competition of anisotropy and adhesion}
\author[itp,scc]{A.Yu.~Menshutin}
\ead{may@itp.ac.ru}
\author[itp,scc]{L.N.~Shchur}
\address[itp]{
Landau Institute for Theoretical Physics, Akademika Semenova av 1-A, Chernogolovka, Moscow Region, 142432, Russia}
\address[scc]{Scientific Center in Chernogolovka, Institutsky 8, Chernogolovka, Moscow Region, 142432, Russia}

\begin{abstract}
Two-dimensional structures grown with Witten and Sander algorithm
are investigated. We analyze clusters grown off-lattice and clusters grown
with antenna method with $N_{fp}=3,4,5,6,7$ and 8 allowed growth directions.
With the help of variable probe particles technique we measure fractal dimension of such clusters $D(N)$
as a function of their size $N$. We propose that in the thermodynamic limit of infinite cluster
size the aggregates grown with high degree of anisotropy ($N_{fp}=3,4,5$) tend to
have fractal dimension $D$ equal to 3/2, while off-lattice aggregates and aggregates with
lower anisotropy ($N_{fp}>6$) have $D \approx 1.710$.
Noise-reduction procedure results in the change of universality class
for DLA. For high enough noise-reduction value clusters with  $N_{fp} \ge 6$
have fractal dimension going to $3/2$ when $N\rightarrow\infty$.

\end{abstract}
\maketitle
\section{Introduction}
Understanding critical properties of non-equilibrium
processes~\cite{Goldenfeld} is still a challenging problem of
contemporary statistical physics. Dynamical growth phenomena is of particular interest, both
theoretical and practical.
The model for the two-dimensional growth
(Diffusion Limited Aggregation) was introduced about thirty years ago
by Witten and Sander~\cite{WS} and analytical solution of the model is still a challenging problem.
Structures generated by DLA algorithm look very similar to those found in nature
and society. Among examples of such structures are~\cite{Nature}: ice crystals on window,
mineral dendrites on the surfaces of limestone, colony of
bacteria, nano-size crystals grown on the crystal
surface, monolayer polymer films, interfaces in Hele-Shaw cell,
urban growth, etc. Similarity of such objects leads to the natural questions: are the properties
of all above mentioned various structures the same? Are the clusters generated by
different models identical or not?


Growth structures are characterized by fractal dimension~\cite{Mandelbrot}. According to the universality
concept of critical phenomena the value of fractal dimension is one
of the basic characteristics of universality class (see, f.e., ref.~\cite{BH}).
The technical difficulties in estimation of fractal dimension in DLA model 
are the high level of fluctuations~\cite{MS} and strong finite-size effects.
To overcome these difficulties we use previously developed
approach of fractal dimension measurement~\cite{MSV} based on
varying the size of probe particles for calculating the harmonic measure averages and
then taking the limit of probe particle size going to zero. 
This procedure smooths out oscillating
dependence of effective fractal dimension on cluster size and allows us to estimate
its asymptotic behavior with good accuracy.

It will be shown in section~\ref{section-measurement} that
such procedure in some sense increases effective cluster size. Therefore,
the resulting values of DLA properties calculated as averages over harmonic measure
are closer to the thermodynamic limit.

In this paper we present our study of the off-lattice DLA model with two additional parameters,
the local field which brakes spatial isotropy and the adhesion coefficient. In Section 2 we present 
details of algorithm used to grow off-lattice aggregates, 
it is followed in Section 3 by description of modifications
of the algorithm that allow us to build lattice clusters with off-lattice algorithm.
In Section  4  we provide details on numerical procedure of our variable probe particles
method. In Section 5 we present results of analysis. In Section 6 we propose morphological diagram. 
We conclude paper with discussion in Section 7.

\section{Off-lattice killing-free algorithm}

We use off-lattice killing free algorithm for cluster generation, developed in~\cite{MS,MSV}. 
In this algorithm, particles are hard balls of unit size which move in 2D in random direction
with fixed step length. Since motion is actually off-lattice one can use some tricks to speed up
the algorithm, e.g. particles could jump with large step length if they are far away from the cluster. 
Another trick that helps us to speed up the simulation is the so-called killing-free technique. 
When it happens that particle goes far enough from the cluster, i.e. it is outside of circle 
of radius $R_r$ (the return radius, which is larger than radius of the cluster), we return this 
particle back to the birth circle $R_b$, $R_b<R_r$. Since in 2D random motion there is zero 
probability to escape to infinity, particle will always return to the origin, i.e. any particle 
will eventually collide with the cluster. Therefore, it is physically not correct to kill particles 
that have come out of some circle $R_d$ (the death radius, which is used in traditional DLA algorithm), 
even if $R_d \gg R_b$ since it will introduce distortion in the harmonic measure. 
One can analytically solve Laplace equation and find angle-dependent probability for such 
particles to intersect $R_b$ for the first time at any point. 
Probability for a particle starting its walk outside circle of radius $R_b$ from 
position $(0,R_0)$ to intersect $R_b$ at angle $\phi$ for a first time is given by formula
\begin{equation}
P(\phi)=\frac{x^2-1}{2\pi(x^2-2x\cos\phi+1)}, x=R_0/R_b
\label{eq-probab}
\end{equation}
In contrast to 2D, in higher dimensions there exists the non zero non-return probability 
for the random walker. This
means that computational model of DLA or any similar process should take 
this fact into account (see paper~\cite{Ziff-3d} with discussion of this subject in 3D).

The algorithm we use is as follows~\footnote{We refer to $R$ as the unit length equal to the
radius of particles and to the length in the process of random walk realization.}:

\begin{enumerate}
\item Set seed particle at the origin at position (0,0).
\item Calculate size of the cluster $R_{max}$. Set birth radius value to $R_b=R_{max}+5R$ and
set return radius value to $R_r=(R_b+200R)*1.2$.
\item Create new particle at random position on the birth radius $R_b$.
\item Calculate distance $d$ from current particle position to the cluster. If
$d<5R$ than move the particle in random direction with step length equal to $R$ and check for collisions with
the cluster, otherwise ($d>5R$) move the particle in random direction with step length equal to $d-5R$.
\item If the particle is out of $R_r$ than return it back to $R_b$ with angle-dependent probability $P(\phi)$
given by formula~(\ref{eq-probab}).
\item Repeat steps (4)-(5) until the particle has been attached to the cluster, then start with step (2).
\end{enumerate}

Effective realization of step~(4) of the algorithm requires a lot of effort. First, to find out
whether a new particle has collided with the cluster we check only its adjacent neighbors.
This requires a special lattice which covers all particles and separates them into different cells.
Doing so we check whether the particle collides or not
only with particles from surrounding cells relative to a current
particle position. We choose a cell size of this lattice to be equal to $32R$.
Second, using the lattice also simplifies and speeds up process of calculation 
of distance $d$ needed at 4-th step
of algorithm. We mark all cells in the lattice as occupied or not and calculate $d$ based on only this information.
This gives lower estimate on $d$. Details for this algorithm are presented in~\cite{MSV}.
Third, we found that it is very effective for memory management to have particles being stored in cells.

One should be careful with system memory allocation procedures since memory allocation
for new particles one by one will produce a great overhead and consume up to 5 times more memory.
Each cell of the lattice has different number of particles in it and it is not reasonable to preallocate
memory for all particles in cell because this number is unknown in advance.
To deal with this we have implemented our own memory allocation
routines which are very effective relative to computer memory requirements. 
Our own memory allocation
procedures and  special memory organization allows us to implement cell-based swap memory.
As long as operating system swapping works with memory pages that are much larger than size of a cell with
average number of particles in it, such large memory region will always contain particles that are on the boundary
of the cluster and thus are always accessed by new coming walkers.
Our cell-based swapping saves a single cell to a disk when the cell was not visited by new particles for a long enough time.
Usage of swapping allows us to generate cluster with up to 200 mln particles on a computer
with only 2Gb of RAM. 

\section{Lattice algorithm and effect of local anisotropy}

\begin{figure}[t]
\centering
\includegraphics[width=6cm]{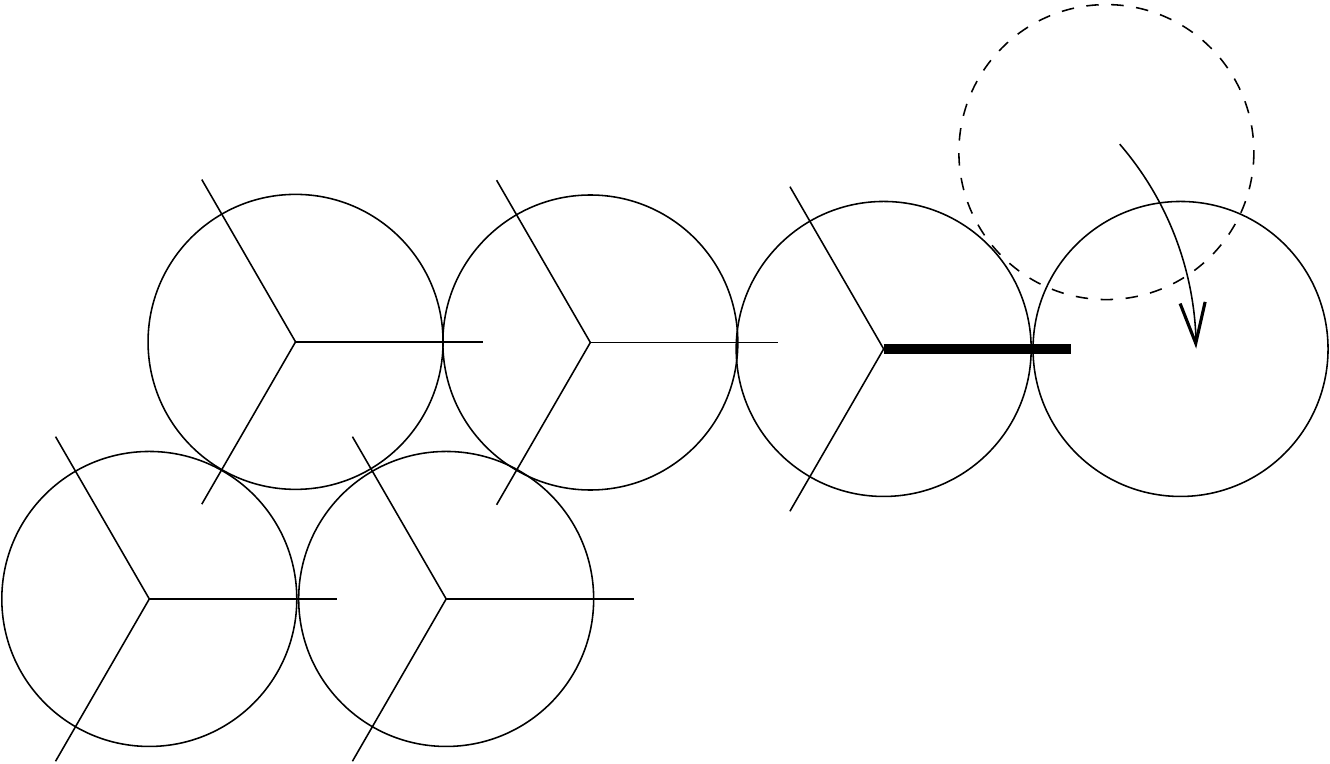}
\caption{Antenna method for new particle attachment. Each new particle (dashed circle) moves to the closest
antenna (bold line) of its parent particle.}
\label{pic-fp}
\end{figure}

Most of the improvements of the algorithm for generation of DLA clusters, described in the previous chapter, are
consequences of its off-lattice nature and they are based on isotropy of random motion in plane. Still there
exists a major class of objects that are either grown on lattice or have some intrinsic anisotropy.
One can imagine a cluster build with atoms that form a lattice. Such a lattice is only being formed
after the atoms have come into contact with each other.
It is possible to model such features without essential changes in our algorithm if one adopts the so
called antenna method~\cite{Meakin-antenna,Eckmann-antenna,Ball-antenna}.
Let particles constituting the cluster be only allowed
to occupy a finite set of positions. This could be imagined as each particle of the cluster has $N_{fp}$ antennas in such a way
that if new particle touches cluster at some point, it is shifted to the closest antenna carried by the
parent particle. All antennas are aligned along
the $N_{fp}$ directions (see Fig.~\ref{pic-fp}). 
For example, the case of $N_{fp}=4$ with anisotropy directions perpendicular to each other
is equivalent to square lattice model. And $N_{fp}=3$ represents hexagonal lattice.
The only difference
with real square lattice model is that the
motion of a particle does not have any anisotropy in it 
until the particle is attached to the cluster.

In our study of DLA with anisotropy in particle attachment we add an extra parameter -- adhesion of particles.
With this property new particles stick to the cluster only after $N_{nr}$ collisions 
with the same antenna, instead of sticking after the first collision. 
Practically, this is the same as
noise reduction in DLA, when each particle of cluster has a counter associated with it.
When new particle hits the cluster, the counter is incremented. The particle is attached to the
cluster only if the counter value is greater than some cut-off value $N_{nr}$, otherwise the particle
is destroyed and new one is released. One can think of the noise-reduction as of growth probability
cut-off which eliminates lowest growth probabilities keeping only the most probable.
It was a long time belief that noise reduction procedure results in faster convergence to asymptotic
properties in DLA~\cite{Nittmann-noise} since it allows only for most probable events to happen.
In our simulation not each cluster particle but each antenna on each particle
has its noise-reduction counter. This
results in unexpected behavior of DLA properties which will be presented later in this paper.

\section{Measurement procedure}
\label{section-measurement}

The basic property of an ensemble of clusters that we are interested in is the cluster fractal dimension $D$. 
It is a common belief that fractal value of cluster dimension should be associated with 
some ``criticality''. 
Such approach is well developed for equilibrium processes and it is known that equilibrium 
criticality is generally described by set of two non-trivial exponents.
In non-equilibrium and dynamical processes identification of criticality is 
still a challenging problem. Therefore, in the present paper
we restrict ourselves with measurement of the value of well defined quantity -- the cluster fractal dimension $D$.

\begin{figure}[t]
\centering
\includegraphics[width=\columnwidth]{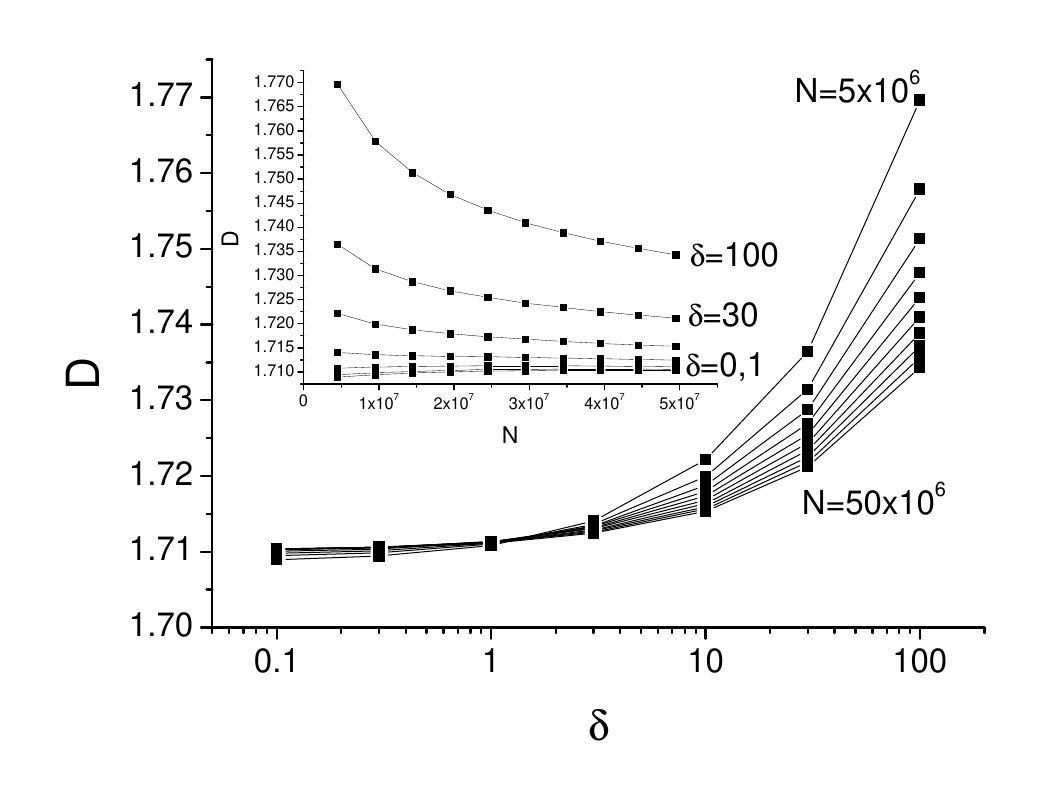}
\caption{Dependence of fractal dimension $D(N,\delta)$ on size $N$ of the off-lattice cluster and on size $\delta$ of probe particles. Calculated over ensemble of 1000 clusters with $5\cdot  10^7$ particles each.}
\label{pic-d_delta}
\end{figure}

Long-time correlations~\cite{MS} and slow vanishing fluctuations~\cite{MSV} make it 
rather difficult to measure fractal dimension $D$ in standard
way and with good precision. To overcome the difficulty, method of variable probe 
particle size was developed in Ref.~\cite{MSV}.
The idea of this method is as follows. 
For each cluster we measure, for example, average deposition radius $R_{dep,i}^{hm}(N,\delta)$ calculated
as harmonic-measure average with probe particles of size $\delta$ at cluster size $N$.
With the help of scaling relation $N\propto {R_{dep,i}^{hm}}^{D_i}$ we extract the fractal dimension  $D_i(N,\delta)$
of each cluster $i$ in ensemble as a function of $N$ and $\delta$.
At the
next step we average $D_i(N,\delta)$ over the ensemble: $D(N,\delta)=1/K\sum D_i(N,\delta)$ (see Fig.~\ref{pic-d_delta}).
And, finally,
we take limit $\delta \rightarrow 0$ and find the quantity $D(N)=D(N,\delta=0)$ we are
interested in.
In order to find the limit $D(N,\delta \rightarrow 0)$ we have used in~\cite{MSV} the power law equation
$D(N,\delta)=D(N)+A{\delta}^{\beta}$. Although this equation fits well experimental
data for $\delta>1$, its precision is not very good for $\delta \rightarrow 0$.
Instead of the power law approximation used in~\cite{MSV}, in this paper we use linear
approximation for $D(N,\delta)$ for $\delta=0.1,0.3,1,3$. We have carried out 
the following check that confirms that the approximation is reasonable. 
For the calculation of $R_{dep}^{hm}$
one needs to average position of probe particles that have touched the cluster at some point.
One can choose to average either position of probe particle center or position of the point of contact.
It is clear that in the limit $\delta \rightarrow 0$ resulting $D(N)$ should not depend
on this choice. We have checked this in our simulations and found that our assumption holds
quite well, i.e. linear approximation $D(N,\delta)=D(N)+A\delta$ could be used
for calculation of $D(N)$.

Taking smaller probe particles gives them higher probability to penetrate deeper into cluster structure inside fjords.
If the fjord structure is similar to rectangular slit of length $l$ and width $a$,
then probability for a particle to pass through it is proportional to
$\exp (-\pi l/a)$. Since particle of finite size $\delta$ affects effective channel
width $a$, using infinitesimal size probe particles will allow greater part of cluster surface to
be exposed thus resulting in higher accuracy of $D(N)$ estimation. Bottoms of fjords are screened strongly
by branches and these regions belong to frozen part of the cluster. Taking them into calculation
with slightly higher probability results in fractal dimension values with better statistical properties.

\begin{figure}[t]
\centering
\includegraphics[width=\columnwidth]{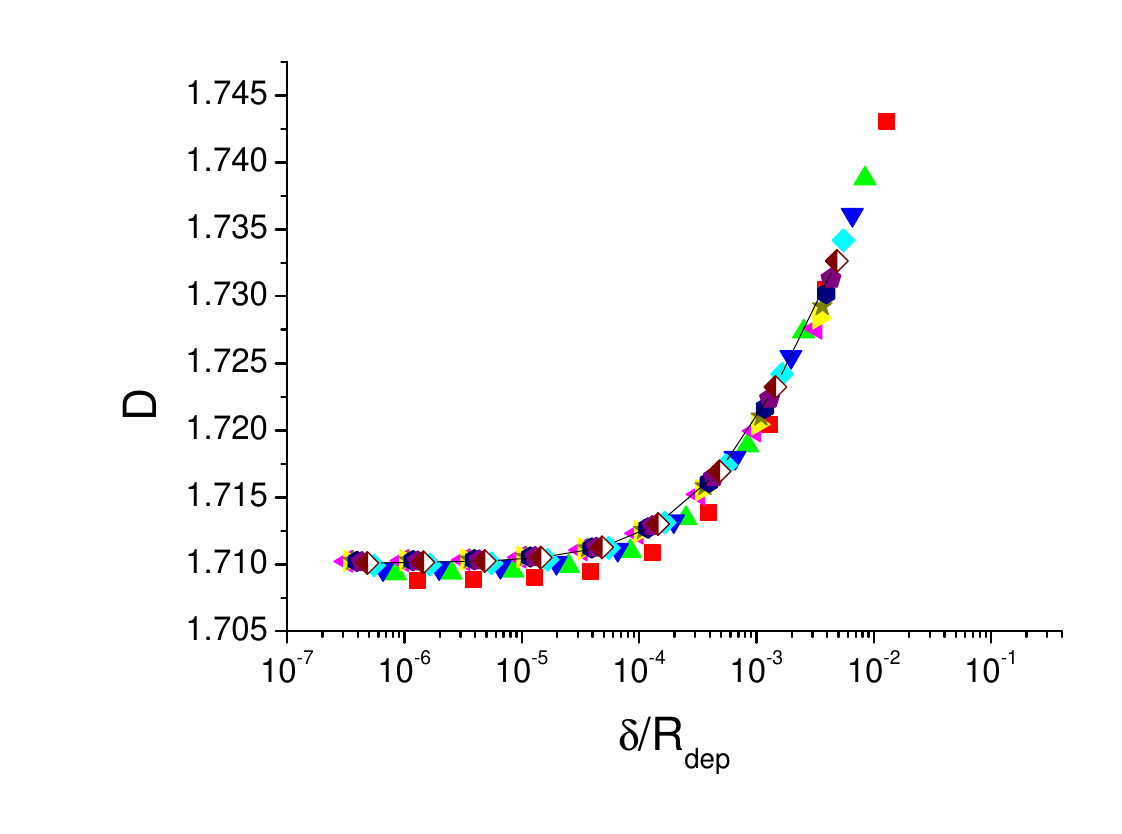}
\caption{(Color online) Rescaled $D(N,\delta)$ dependence (shown in Fig. 2) in the form $D(\delta/R_{dep}(N))$. Different colors represent
different cluster sizes $N$.}
\label{pic-scaling}
\end{figure}

While we cannot treat the effect of variable probe particles analytically to explain
that $D(N)$ obtained with the described procedure is a good quantity  and extrapolates
well the asymptotic fractal dimension of DLA,
we propose the following empirical arguments.
It is easy to see that curves $D(N,\delta)$ in Fig.~\ref{pic-d_delta} are very similar, and
it is actually possible to merge them into a single curve.
For an infinitely large cluster, i.e. for $N=\infty$, the finite size
of particles would be unimportant. But for finite cluster size we should take size
of particles into account. In addition to the size of the cluster there exists only one more natural
length scale -- the particle size. Although we cannot change size of particles in cluster because
all lengths are proportional to this quantity, we can change size of particles during
the measurement procedure. This reasoning leads us to the idea to rescale $\delta$ parameter 
in $D(N,\delta)$ dependence in the following form: $D(\delta/R_{dep}^{hm}(N))$.
Surprisingly, after this transformation
all the curves corresponding to different cluster sizes merge into a single curve, 
as seen in Fig.~\ref{pic-scaling}.
It is also natural to rescale the size of probe particle in this form since the probability for a particle
to pass through the fjord is proportional to the difference in fjord width and particle size,
and the width of the fjord is directly proportional to the $R_{dep}$, if we assume that the number of branches does
not change with the time. In other words, there are only two independent length scales $\delta$ and $R_{dep}$.
The existence of such scaling means that if we take probe particles 
ten times smaller than usual the fractal dimension will be the same as for the cluster
that has $R_{dep}$ ten times bigger and is measured with probe particles
of unit size. Thereby, measuring fractal dimension of DLA in the limit of probe particles
of vanishing size results in values converging faster to the asymptotic thermodynamic regime.

\section{Results}

We have generated clusters with different number $N_{fp}$ of axes of the anisotropy field,
varying from $N_{fp}=3$ to 8, as well as off-lattice clusters
(this case is equivalent to $N_{fp}\rightarrow\infty$).
Each ensemble has 1000 clusters and each cluster consists of 50 millions of particles.
\begin{figure}[t]
\centering
\includegraphics[width=\columnwidth]{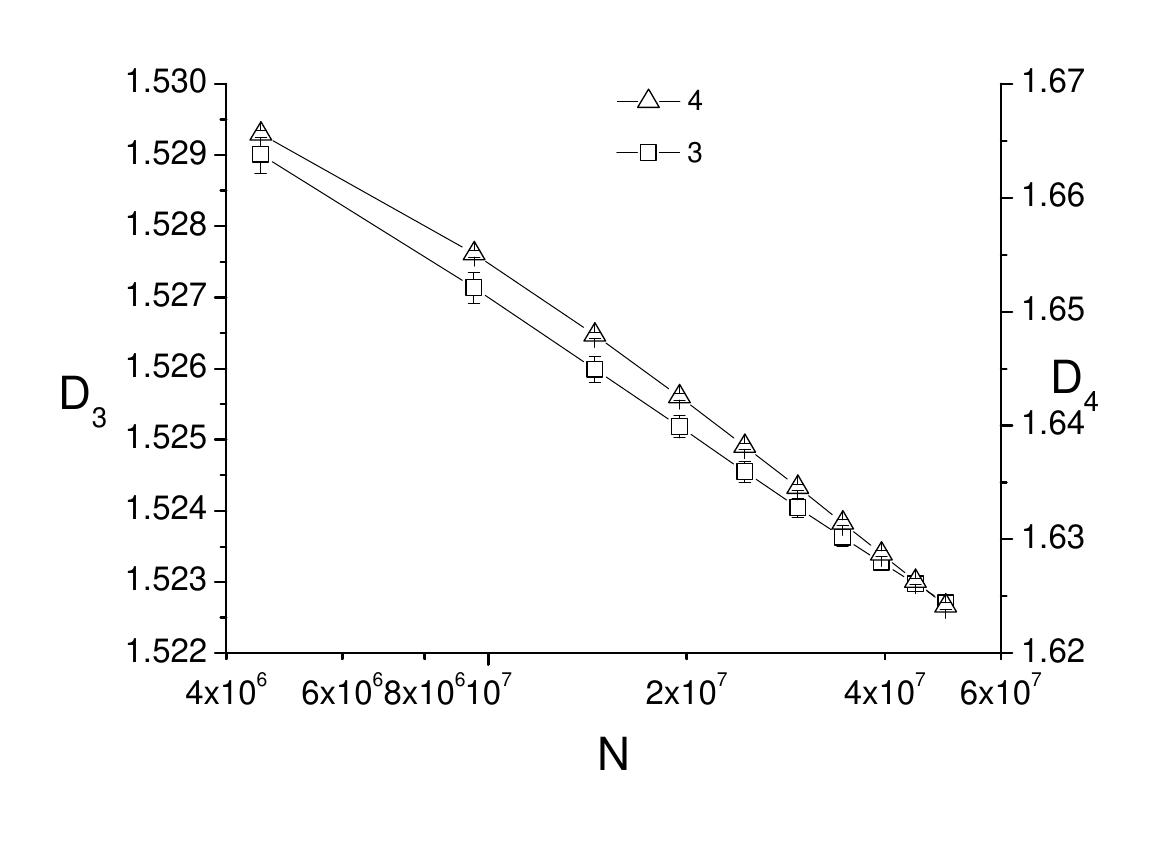}
\caption{Fractal dimension $D(N)$ of clusters grown with $N_{fp}=3$ ($D_3$, left axis) and 4 ($D_4$, right axis)
anisotropy axes as a function of cluster size $N$.}
\label{pic-d-3-4}
\end{figure}

\begin{figure}[t]
\centering
\includegraphics[width=\columnwidth]{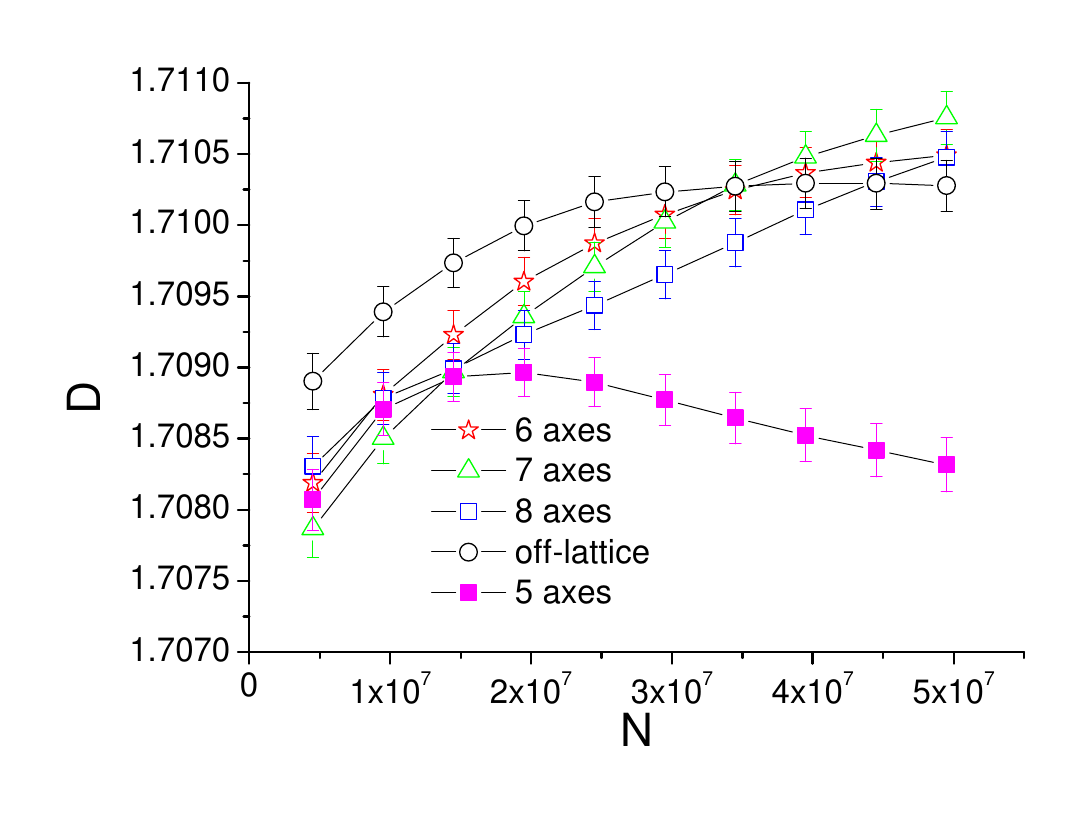}
\caption{(Color online) Fractal dimension $D(N)$ of clusters grown with $N_{fp}=5,6,7,8$ anisotropy axes and also
of off-lattice clusters as a function of cluster size $N$.}
\label{pic-d-5-off}
\end{figure}



Fractal dimension of the clusters, calculated with the help of technique described in Sec.~\ref{section-measurement},
is presented in Figs.~\ref{pic-d-3-4} and ~\ref{pic-d-5-off}.
Fractal dimension $D(N)$ for clusters with 3 axes (hexagonal lattice) diminishes 
with increasing the size of the cluster as $D_3(N)=1.5699(14)-0.00613(19)\log_{10}N$;
for clusters with 4 axes (square lattice) $D_4(N)=1.9400(22)-0.04093(30)\log_{10}N$.
Therefore, for big enough clusters fractal dimension of on-lattice models goes down.
According to prediction of Kesten, the lowest possible fractal dimension
for such objects is exactly $3/2$~\cite{Kesten}. However, to reach this value one needs to generate cluster
with at least $10^{11}$ particles. This value is 3 orders of magnitude higher
than value reachable with modern computers and algorithms.

In contrast to the previous case, fractals with $N_{fp}>5$ tend to have fractal dimension
$D$ equal to dimension of off-lattice clusters. The closer the branches of DLA are located,
the higher are screening effects and the higher are fluctuations that destroy anisotropy.
\begin{figure}[t]
\centering
\includegraphics[width=\columnwidth]{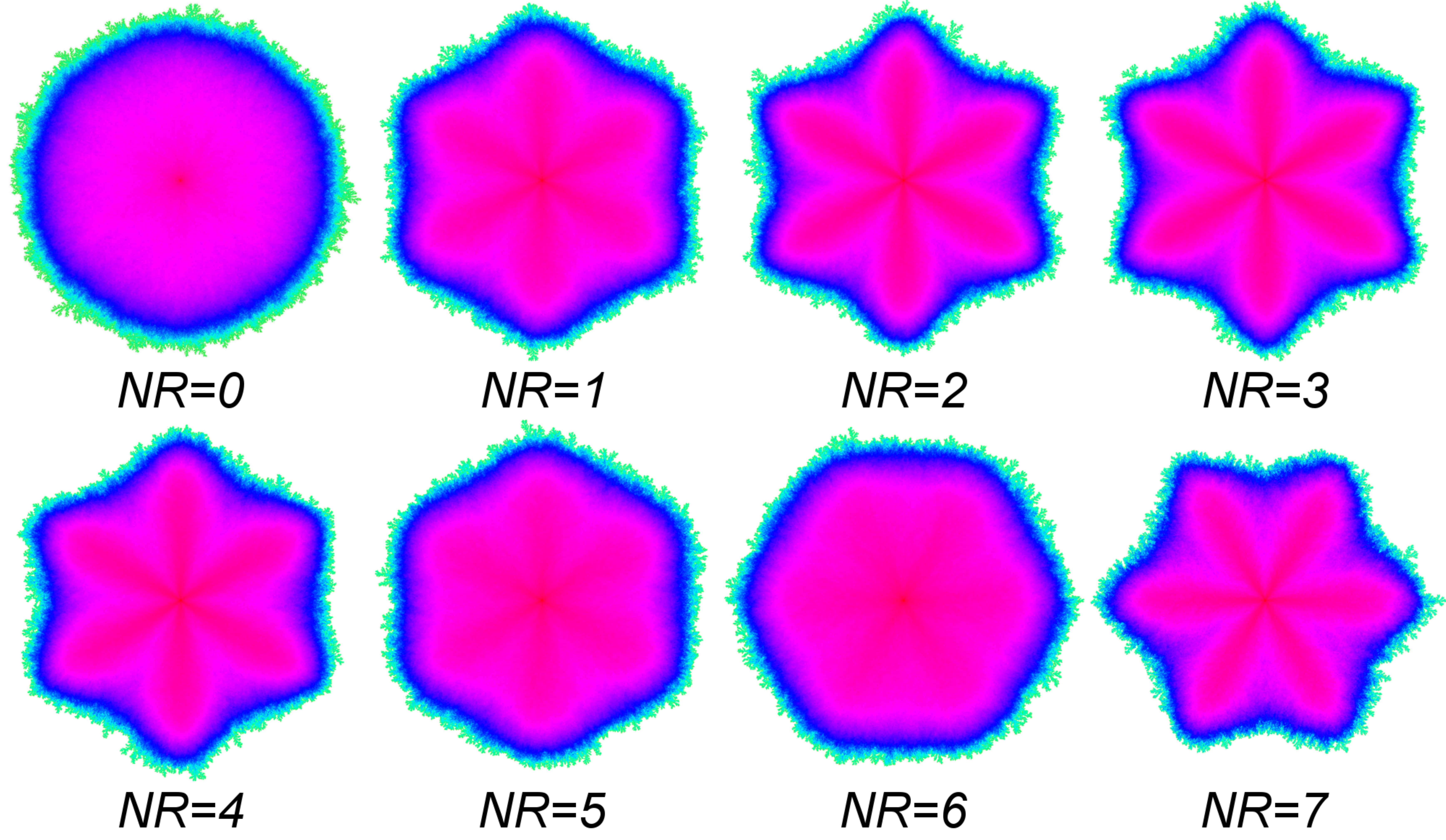}
\caption{(Color online) Average density of particles in ensemble of clusters grown with six
anisotropy axes with varying noise reduction level $N_{nr}$=0,1,2,3,4,5,6,7.}
\label{pic-density-fp6-nr}
\end{figure}
\begin{figure}[t]
\centering
\includegraphics[width=\columnwidth]{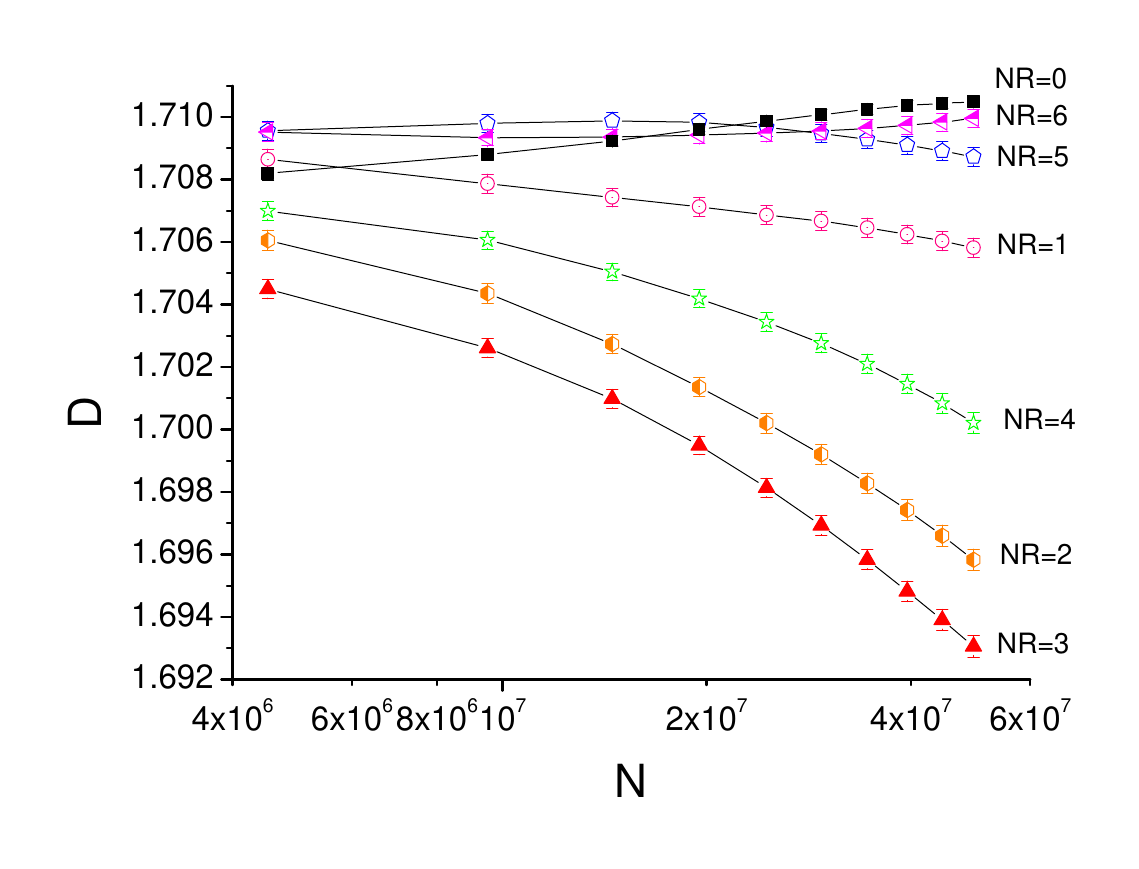}
\caption{(Color online) Fractal dimension $D(N)$ of clusters grown with $N_{fp}=3$ anisotropy axes and varying noise reduction level as a function of cluster size $N$.}
\label{pic-d-fp6-rn}
\end{figure}

Change in $N_{fp}$ anisotropy parameter not only influences fractal dimension of the
cluster but also results in transition in the geometrical structure of DLA.
Clusters with $N_{fp}=3,4$ (hexagonal and square lattices) prefer to grow along
anisotropy axes of the cluster. The number of main branches for such clusters
is equal to $N_{fp}$. Clusters with $N_{fp}>5$ look isotropic on the average and
do not have any preferred growth directions. Peculiar case $N_{fp}=5$ separates
these two well-defined regions.  Asymptotic behavior in that case
is obscured by slow transition effects. Looking at Fig.~\ref{pic-d-5-off}
one may suggest that in the limit $N\rightarrow\infty$ 
the asymptotic value of the fractal dimension $D_5$ will be $3/2$. 
It is worthwhile to mention that
on the average clusters with $N_{fp}=5$ also prefer to grow along several preferred directions,
as in $N_{fp}=3,4$ case. However, in contrast to the case $N_{fp}=3,4$,
these directions are not alongside with anisotropy axes but
are along the axes turned by the angle $\pi/(2 N_{fp})$, i.e.
they are strictly between the anisotropy axes.

We also analyze influence of the noise reduction on the structure
of the cluster. Let us start first with $N_{fp}=6$ case. Average density of particles
for clusters with six anisotropy axes is presented in Fig.~\ref{pic-density-fp6-nr}, where 
average was taken over ensemble of 1000 clusters.
The parameter $N_{nr}$ represents the noise reduction level 
used to grow an ensemble of clusters. The case $N_{nr}=0$ corresponds
to the usual simulation without noise reduction,  $N_{nr}=1$ means that particle
will be attached on the second collision with the particular antenna, etc.
Clusters without noise reduction are isotropic and symmetric. 
Anisotropy decrease becomes more pronounced with increasing the noise reduction, 
however preferred growth directions
of cluster are no more aligned with the antenna directions~\footnote{The first antenna
(anisotropy axis) is always aligned along X axis in all figures.}.
Further increase of $N_{nr}$ makes cluster more asymmetric and at $N_{nr}=6$ main branches of the cluster
are being rotated by the angle $\pi/(2 N_{fp})$.

Alteration in geometrical structure is also reflected on fractal dimension of the system.
This could be seen from the Fig.~\ref{pic-density-fp6-nr}. As soon as the cluster becomes
anisotropic, its fractal dimension decreases until $N_{nr}=3$. After that its structure
tends to be more symmetric ($N_{nr}=5$) until rapid change in
preferred growth directions at $N_{nr}=6$ occurs. Further increase in value of
$N_{nr}$ leads to decrease in $D(N)$ with size of the cluster while aggregate structure
always stays asymmetric and its branches grow along anisotropy directions.

\section{Morphological diagram and phase transition }

\begin{figure}[t]
\centering
\includegraphics[width=\columnwidth]{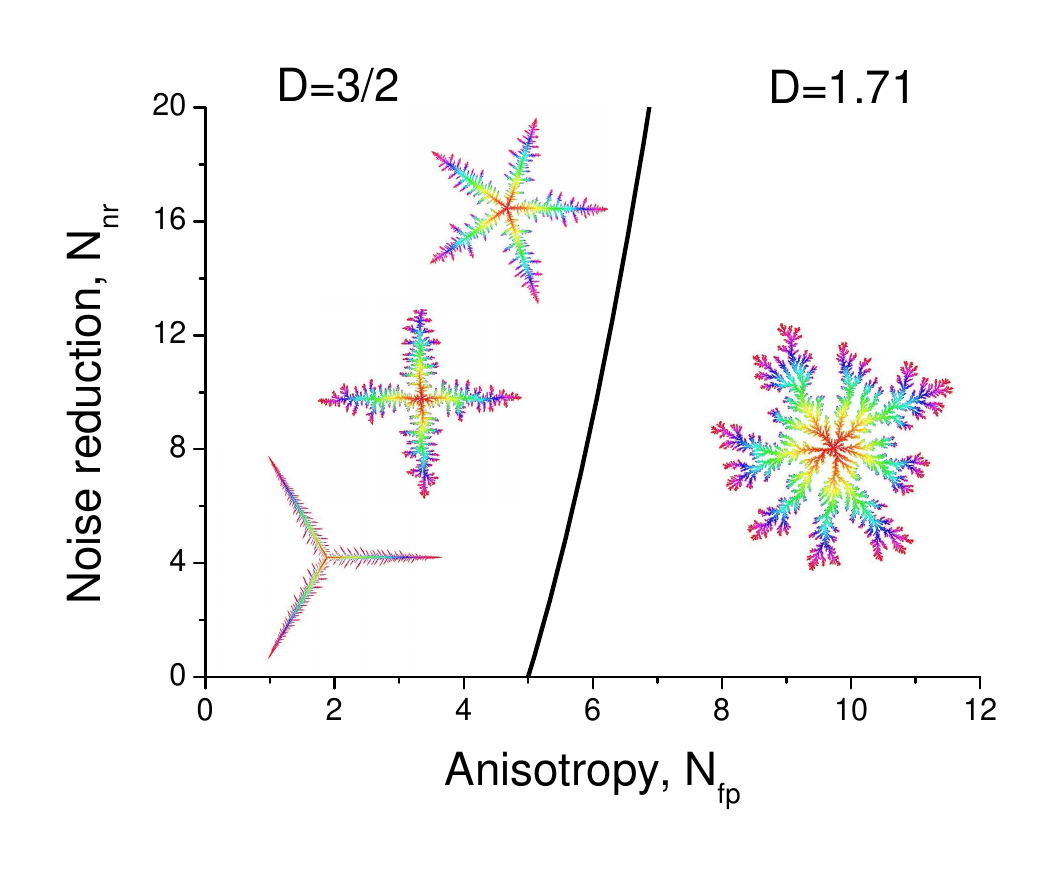}
\caption{(Color online) Morphological diagram of DLA clusters. The region left from the solid line corresponds to the $N_{fp}$-fold 
fractal crystals with asymptotical fractal dimension 3/2. The region to the right of the solid line corresponds to the random fractal with asymptotic fractal dimension 1.710.
Point $(N_{fp}=5,N{nr}=0)$ is marginal point.}
\label{pic-phase-diagram}
\end{figure}

Based on the properties of clusters presented in the previous section we propose
the following picture that separates all two-dimensional DLA-like objects
into two major classes. Clusters with low number of anisotropy growth directions
$N_{fp}<6$ and clusters with large number of growth direction but grown with
noise-reduction procedure $N_{nr}>N_{nr}^{crit}(N_{fp})$ have anisotropic
shape with number of main branches equal to $N_{fp}$. Their fractal dimension
is $3/2$ in the thermodynamic limit of big enough systems.
Another class of objects consists of structures that look symmetrical on the average
with no well-defined preferred direction. Their fractal dimension approaches to
the well known value of $1.710$ corresponding to off-lattice DLA clusters. We sketch this result in the
diagram shown in Fig.~\ref{pic-phase-diagram}.

\section{Results and discussion}
Based on the intensive numerical simulations 
we propose morphological diagram for two-dimensional structures which may be typical 
for all processes where diffusion dominates growth.
At first glance similar behavior was found in paper by Nittmann and Stanley~\cite{Nittmann-noise} for six-fold
symmetry structures grown with DBM algorithm. But in contrast to~\cite{Nittmann-noise} 
our simulation for clusters with $N_{fp}=3,4,5$ results in fractal dimension equal to $1.5$, while Nittmann and Stanley
propose that it is equal to $1.71$. Also finite noise reduction in our simulation could
change structure of the aggregate from one class to another in contrast to~\cite{Nittmann-noise}.
In our model anisotropy is not required to be specially enhanced to be noticeable.

In the paper by Aukrust and others~\cite{Aukrust} authors have also proposed morphological
transition taking place when the level of anisotropy was varied. But authors have studied only square-lattice
model where anisotropy was varied with number of local neighbors.

In our simulation we have focused on effects caused by underlying lattice anisotropy for
different lattices ($N_{fp}=3,4$) and pseudo-lattices $N_{fp}>5$. Since $N_{fp}=5$ is the marginal
case that separates $N_{fp}$-fold clusters and fractal clusters, we also propose that the number of
main branches in DLA is somewhere between 5 and 6. This number is slightly higher than $4.8$ 
predicted by Ball~\cite{Ball-antenna} but is still very close to that value.

A. Yu. Menshutin thanks Dynasty Foundation for supporting this work.


\end{document}